\documentclass[11pt,twoside]{article}
\usepackage{psfig}
\usepackage{jltp}

\newcommand{\srruo}{Sr$_2$RuO$_4$}

\newcommand{\srdzs}{Sr$_3$Ru$_2$O$_7$}

\newcommand{\caruo}{Ca$_2$RuO$_4$}

\newcommand{\csruo}{Ca$_{2-x}$Sr$_{x}$RuO$_4$}
\newcommand{\caruoea}{Ca$_{1.8}$Sr$_{0.2}$RuO$_4$}
\newcommand{\caruoef}{Ca$_{1.5}$Sr$_{0.5}$RuO$_4$}

\newcommand{\casrruo}{Ca$_{2-x}$Sr$_{x}$RuO$_4$}
\newcommand{\muB}{$\mu_B$}

\newcommand{\gbd}{$\gamma$-band}

\newcommand{\bbd}{$\beta$-band}

\newcommand{\dxy}{$d_{xy}$}
\newcommand{\dxz}{$d_{xz}$}
\newcommand{\dyz}{$d_{yz}$}

\newcommand {\vq}{$\mathbf{q}$}

\title{Magnetoelastic coupling across the metamagnetic
transition in Ca$_{2-x}$Sr$_x$RuO$_4$ ($0.2 \leq x \leq 0.5$)}

\author{J.~Baier,$^1$ P.~Steffens,$^1$
O.~Schumann,$^1$ M.~Kriener,$^{1,2}$ S.~Stark,$^1$ H.~Hartmann,$^1$
O.~Friedt,$^1$ A.~Revcolevschi,$^3$ P.G.~Radaelli,$^4$
S.~Nakatsuji,$^2$ Y.~Maeno,$^2$ J.A.~Mydosh,$^1$ T.~Lorenz,$^1$ and
M.~Braden$^1$}
\address{$^1$II.\,Physikalisches Institut, University of Cologne,
Z\"ulpicher Str. 77, 50937 K\"oln, Germany \\
e-mail : braden@ph2.uni-koeln.de \\
 $^2$\,Department of Physics, Kyoto University, Kyoto 606-8502,
 Japan\\
 $^3$\,Lab. de Physico-Chimie de  l'\'Etat Solide, Universit\'e Paris-Sud, 91405 Orsay Cedex, France \\
$^4$\,ISIS Facility, Rutherford Appleton Laboratory-CCLRC,
Chilton, Didcot, Oxfordshire OX11 0QX, United Kingdom}

\runninghead{\bf J.~Baier {\boldmath $et\ al.$}}{Magnetoelastic
coupling in Ca$_{2-x}$Sr$_x$RuO$_4$ ($0.2 \leq x \leq 0.5$)}

\begin{document}

\begin{abstract}

The magnetoelastic coupling in \caruoea \ and in \caruoef \ has
been studied combining high-resolution dilatometer and
diffraction techniques. Both compounds exhibit strong anomalies
in the thermal-expansion coefficient at zero and at high magnetic
field as well as an exceptionally large magnetostriction.  All
these structural effects, which are strongest in \caruoea , point
to a redistribution of electrons between the different $t_{2g}$
orbitals tuned by temperature and magnetic field. The temperature
and the field dependence of the thermal-expansion anomalies in
\caruoea \ yield evidence for a critical end-point lying close to
the low-temperature metamagnetic transition; however, the
expected scaling relations are not well fulfilled.

\end{abstract}

\maketitle


\section{INTRODUCTION}

The phase diagram of \csruo \ exhibits a rich variety of physical
phenomena spanning the unconventional superconductor \srruo \ to
the antiferromagnetic Mott-insulator \caruo \cite{1,2,3,4}. The
essentially different character of these physical ground states
is quite outstanding in view of the fact that only the ionic
radius on the Ca/Sr-site varies throughout this series. The
\casrruo -phase diagram offers therefore the interesting
possibility to tune through a Mott-transition by structural
changes only. The smaller ionic radius of divalent Ca compared to
that of divalent Sr induces a series of structural phase
transitions characterized by rotations or tilts  of the
RuO$_6$-octahedra as it is typically observed in perovskites and
related compounds. Such structural deformations have a strong
impact on the electronic band structure since they modify the
metal-oxygen hopping parameters and thereby the electronic band
widths \cite{5,6}. In \casrruo \ the decrease of the Sr content
$x$ first stabilizes a rotation of the octahedra around the $c$
axis and then, for $x \leq 0.5$, a tilting of the octahedra
around an in-plane axis \cite{7}. Further decrease of the
Sr-content finally leads to the Mott-transition associated with
another structural transition, across which the RuO$_6$ octahedra
become flattened and their tilting increases \cite{7,8,9}.

Remarkable physical properties were reported for the
Sr-concentration range $0.2 \leq x \leq 0.5$, i.e. in the
metallic phase close to the Mott transition \cite{8,9}.
Approaching the Sr content x=0.5 from higher values Nakatsuji et
al. report a continuous increase of the low-temperature magnetic
susceptibility reaching at x=0.5 a value 200 times larger than
that of pure \srruo \cite{8,9}. Furthermore, the electronic
coefficient of the specific heat is exceptionally high, of the
order of ${C_p/T}\sim 250\,\frac{\rm mJ}{\rm mole\,K^2}$
\cite{8,9,jin}, well in the range of typical heavy fermion
compounds. Inelastic neutron scattering has revealed strongly
enhanced magnetic fluctuations  \cite{friedt-prl} with a
propagation vector of \vq $\sim$(0.2,0,0). The fluctuations in
\casrruo \ with $x$ close to 0.5 are quite different from those
in pure \srruo\ \cite{sidis,braden2002}: Although the magnetic
instability observed for $x=0.5$ is still incommensurate, its
character is closer to ferromagnetism. The magnetic properties of
the \casrruo -compounds with a Sr content close to 0.5 show some
resemblance to localized electron systems; it has even been
proposed that in these materials an orbital-selective
Mott-transition occurs leaving a part of the 4d-electrons
itinerant \cite{10}. This proposal has initiated a strong debate
concerning its theoretical basis as well as concerning its
applicability to the phase diagram of \csruo . It appears
 safe to assume however, that the \gbd \ associated with the \dxy
-orbital exhibits a much smaller band width than that of the
$\alpha$- or \bbd , since the rotation mainly influences the
hybridization of the \dxy -electrons \cite{5,6}.

Upon further decrease of the Sr-content the tilt transition
occurs with an apparently strong impact on the magnetic
properties. The low-temperature magnetic susceptibility rapidly
decreases with increasing tilt and the electronic specific-heat
coefficient is reduced but remains at a rather high level.
Applying a magnetic field to \caruoea \ at low temperature induces
a metamagnetic transition with a step-like increase of the
magnetisation of about 0.4\ \muB \ per Ru. The metamagnetic
transition field, $H_{mm}$  sensitively depends on the direction
of applied field; the transition occurs at 5.5\ T when the field
is applied along the $c$-direction \cite{9} whereas values of 2
and 7\ T are found for field directions along the $a,b$-plane
\cite{balicas}. The strong anisotropy of the metamagnetic
transition field, in particular the difference for the two
orthorhombic in-plane directions, suggests the relevance of spin
orbit coupling. The meta\-magnetic transition in \caruoea \
strongly resembles that observed in the double-layer ruthenate
\srdzs , which has attracted special interest due to
quantum-critical phenomena related with the end-point of the
first-order metamagnetic transition\cite{11,12,13}. Apart from
the different basic structure, these two- and single-layer
ruthenates possess similar structural characteristics, in
particular, they both exhibit the structural deformation
characterized by octahedra rotation around the $c$-axis.

In recent work, we have analyzed the structural aspects of the
metamagnetism in \casrruo \ by a combination of diffraction, of
high-resolution thermal expansion, and of magnetostriction
experiments \cite{kriener,baier-physb}. These studies gave
evidence of a change of the orbital arrangement driven by either
temperature or magnetic field. The thermal-expansion anomalies
observed at zero magnetic field illustrate an increase of the \dxy
\ orbital occupation upon cooling, whereas the anomalies at high
field point to a decrease of the \dxy-occupation upon cooling.
Accordingly, the structural effects seen as a function of magnetic
field at low temperature (diffraction and magnetostriction
results) show that upon increasing magnetic field electrons are
transferred  from the \dxz - and \dyz -orbitals into the \dxy
-orbital. In this work we have completed these studies for
\casrruo \ with $x=0.2$ and $x=0.5$ by additional diffraction
studies and by high-resolution dilatometer measurements along
different direction in longitudinal and transverse magnetic
fields. In addition we have focused on the metamagnetic
transition itself by collecting more data close to the critical
field and by extending the measurements towards lower
temperatures for \caruoea \ where the metamagnetic transition is
best defined.

\section{EXPERIMENT}

Single crystals of Ca$_{2-x}$Sr$_x$RuO$_4$ were grown by a
floating-zone technique in image furnaces at Kyoto University $(x
= 0.2)$  and at Universit\'e Paris Sud $(x = 0.5)$. Details of the
preparation process are reported in reference ~\cite{nakatsuji01a}
. In addition, powder samples of Ca$_{1.8}$Sr$_{0.2}$RuO$_4$ and
\caruoef \ were prepared following the standard solid-state
reaction. The samples were from the same batches as those studied
in Refs.~\cite{8,9} or characterized by x-ray diffraction and
susceptibility to possess identical properties.

Thermal expansion and magnetostriction were studied in magnetic
fields up to 14 T in two different home-built capacitive
dilatometers down to a lowest temperature of 300\,mK
\cite{braendli73a,lorenz97a,pott83a,heyer}. The magnetization
measurements were performed using a Quantum Design vibrating
sample magnetometer (VSM).  With the GEM diffractometer at the
ISIS facility, neutron powder diffraction patterns were recorded
as a function of temperature and in fields up to 10 T for
\caruoea. X-ray powder diffraction patterns were recorded between
10 and 1000K using a D5000 Siemens diffractometer and
Cu-K$_{\alpha}$-radiation.

\begin{figure}
\centerline{\psfig{file=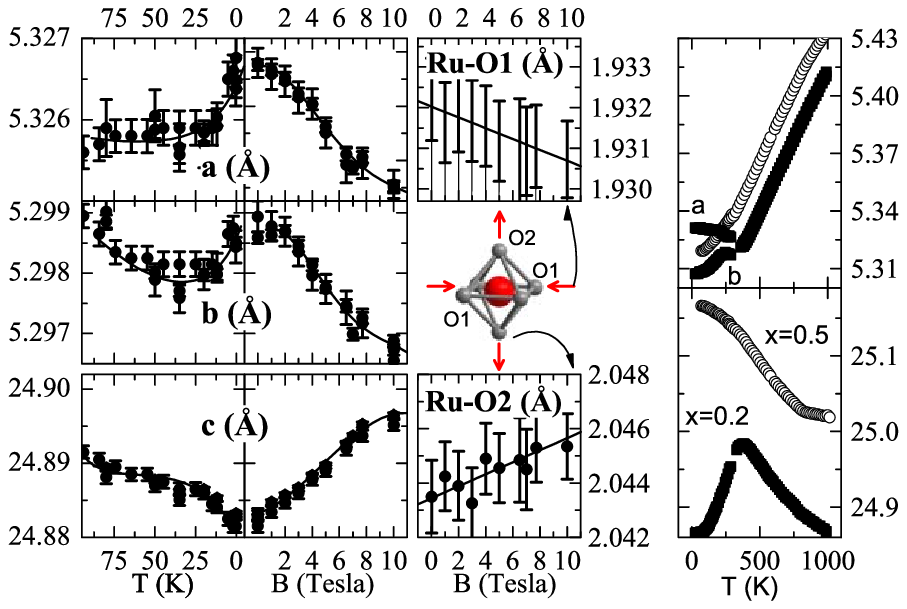,width=11cm}} \caption{Results
of x-ray and neutron diffraction studies on \caruoea \ and
\caruoef. The left column gives the three orthorhombic lattice
constants in \caruoea \ as a function of temperature (results
obtained on GEM); the middle columns show the lattice constants
and the RuO bond lengths in \caruoea \ as  a function of magnetic
field. The right column shows the $a$ and $b$ parameters (above)
and the $c$ lattice constant (below)  for a wider temperature
range for x=0.2 and 0.5. Note that \caruoea  \ exhibits a
tetragonal to orthorhombic transition above room temperature
whereas \caruoef \ stays tetragonal down to about 50\ K.
}\label{diffraction}
\end{figure}

\section{RESULTS AND DISCUSSION}

\subsection{Metamagnetic transition in \caruoea }

Among the \casrruo \ series the metamagnetic transition is best
defined in \caruoea , therefore, we have chosen this composition
for the most detailed thermodynamic studies. Figure 1 presents
the results of diffraction experiments on a powder sample to
characterize the structural evolution as function of temperature
and magnetic field. These data were taken with the GEM
diffractometer using a magnetic-field cryostat. In addition, the
right panels of  Figure 1 show the lattice parameters determined
by x-ray diffraction in a wider temperature interval. \caruoea
exhibits two distinct structural distortions at low temperatures,
the rotation of the RuO$_6$-octahedrons around the c-axis and the
tilting. Above $\sim$350\ K, we find only the rotational
distortion in space group $I4_1/acd$ with a lattice of
$\sqrt{2}\cdot a, \sqrt{2} \cdot a, 2\cdot c$ with respect to the
ideal tetragonal lattice. A characteristic feature of this phase
observed in many \casrruo -compounds \cite{steffens-unp} concerns
the negative thermal expansion along the c-axis which persists
over a broad temperature range, see the x-ray data in Fig. 1. The
structural phase transition associated with the octahedra tilting
further reduces the symmetry towards $Pbca$ with the nearly same
lattice parameters as in $I4_1/acd$. Although the $Pbca$ space
group has also been reported for the insulating and metallic
phases observed for $x < 0.2$ , see reference \cite{7}, the
symmetries are different, since in the case of the
low-temperature phase for $x \geq 0.2$ the rotational distortion
still leads to a doubling of the $c$-axis. High-resolution
measurements of the thermal expansion along and perpendicular to
the $c$-axis reveal strong low-temperature anomalies
\cite{kriener}, which are also visible in the diffraction data.
Both in-plane parameters expand upon cooling whereas the c-axis
shrinks, see figure 1. The diffraction data further show that upon
increase of the magnetic field both in-plane directions shorten
while $c$ elongates. With the full structure analysis one may
attribute this effect essentially to a change in the octahedron
shape, which becomes elongated at high field indicating a shift of
orbital occupation from \dxy \ to \dxz \ and \dyz \ states
\cite{kriener}.

A set of powder-diffraction patterns were recorded on GEM using a
zero-field cryostat  in order to better characterize the
temperature dependence of the crystal structure in a wider range.
Up to a temperature of 160\ K the essential structural change
arises from a weak variation of the tilt distortion. At low
temperature the tilt angle saturates at values of 5.9 and 4.4
degrees determined at the basal (O1) and apical oxygen (O2),
respectively. The minor differences in these two tilt angles
indicate that the RuO$_6$-octahedrons are not perfect in \caruoea
. The tilt distortion is coupled to the orthorhombic strain $a >
b$ which at first sight is counterintuitive, as the tilt occurs
around the $b$-axis which is the shorter one. Similar to other
K$_2$NiF$_4$ materials with a tilt distortion, the interactions
in the Ca/Sr-O rock-salt layer induce an elongation of the
lattice and of the RuO$_6$-octahedron perpendicular to the tilt
axis. Although this octahedral distortion might be relevant in
splitting the \dxz \ and \dyz -$t_{2g}$-levels and hence cause
the in-plane anisotropy of the metamagnetic transition field, it
is not related to an orbital ordering effect. The orthorhombic
splitting, as well as the difference in the O1-O1-edge lengths of
the octahedrons, are clearly coupled to the tilt angles, see Fig.
1 and 2. The low-temperature anomalies seen in the thermal
expansion \cite{kriener} are clearly visible in the lattice
parameters, but the effect in the internal crystal structure is
within the error of this measurement, see Fig. 2.

\begin{figure}
\centerline{\psfig{file=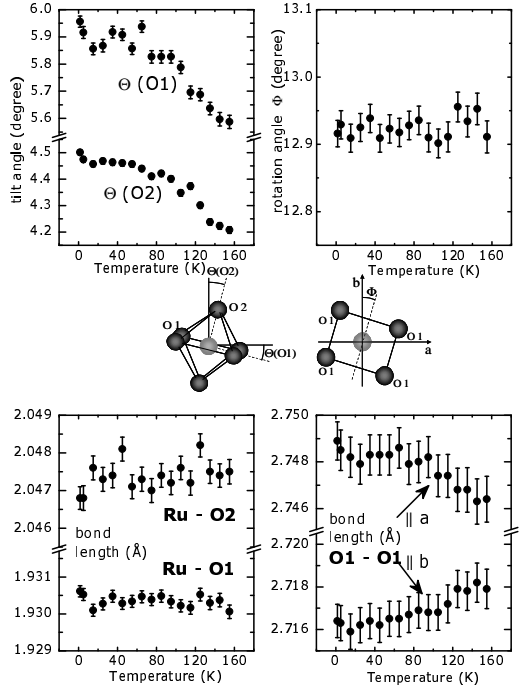,width=6.8cm}}
\caption{Structural evolution of \caruoea \ as determined by
powder neutron diffraction on GEM without magnetic field, the
structural results correspond to Rietveld-fits in space-group
$Pbca$, but constraints had to be used to limit the number of
free parameters.} \label{diffraction-b}
\end{figure}

Figure 3 shows the magnetostriction data recorded along the c-axis
for field directions parallel and perpendicular to $c$.
Qualitatively, both orientations show the same effect, -- the
elongation of the lattice when passing the metamagnetic
transition  at $H_{meta-magn}$=5.7\ T for the field applied along
$c$ and at 2.0\ T for the field perpendicular to $c$,  see also
reference \cite{kriener}. At fields well above the metamagnetic
transition, there is little quantitative difference between the
two field directions: the elongation is only 20\% smaller when
the field is applied parallel to $c$. The comparable structural
effects along both directions imply that the magnetostriction of
\caruoea \ does not simply arise from the alignment of
anisotropic ionic coordinations as it is the case in rare-earth
compounds. Instead it has to be attributed to a transition or at
least to a cross-over between distinct phases. Spin-orbit
coupling seems to cause the sizeable anisotropy of the
metamagnetic transition field, but it seems not to play a major
role in the metamagnetic transition itself. The magnetostriction
data and their field derivative show that the transition becomes
smeared out with increasing temperatures. The maxima of the
derivatives shift towards higher fields with increasing
temperatures. Furthermore, the height of the peak in the
magnetostriction rapidly decreases with increasing temperature
for both field directions; roughly $1 \over \lambda _{max}$
scales with $(T^2 + const.)$, see Figure 2e) and f).

\begin{figure}
\centerline{\psfig{file=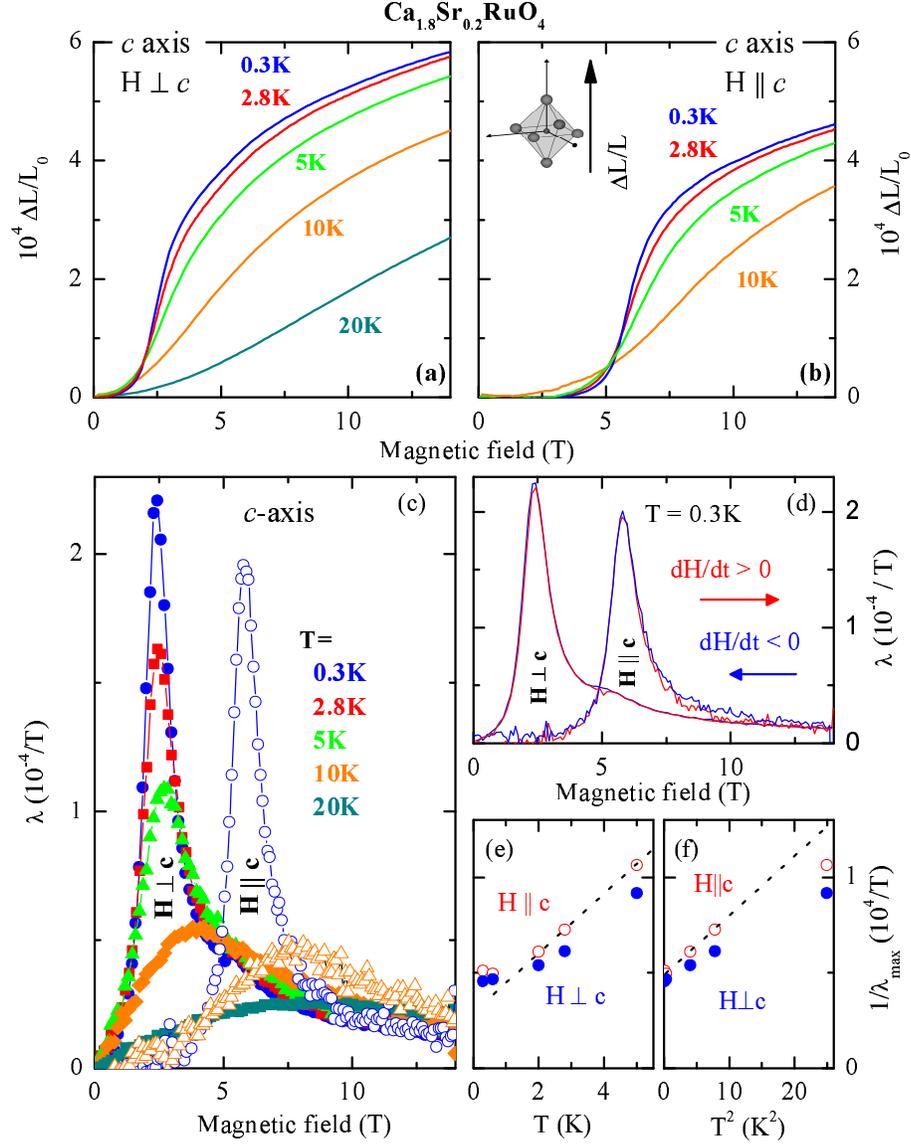,width=12cm}}
\caption{Magnetostriction $\Delta L(H)/L_0$ along the $c$ axis of
Ca$_{1.8}$Sr$_{0.2}$RuO$_4$, measured in a field applied parallel
(a) and perpendicular (b) to the $c$ axis. Panel (c) displays the
derivative $\lambda(H)=\frac{1}{L_0}\frac{\partial L}{\partial H}$
for the $c$ axis in a magnetic field from selected measurements
with d$H$/d$t > 0$. A comparison of $\lambda(H)$ for increasing
(d$H$/d$t
> 0$) and decreasing (d$H$/d$t < 0$) field is presented in panel (d).
The inverse peak height $1/\lambda_{\rm max}$ of the $\lambda(H)$
anomaly at the metamagnetic transition is plotted versus $T$ and
$T^2$ in panel (e) and (f), respectively.} \label{fig:baier_fig1}
\end{figure}

At the lowest temperature of 0.3\ K, the magnetostriction along
both field directions was measured upon increasing and decreasing
field. There is no hysteresis discernible, which is surprising in
view of the expected first-order character of the metamagnetic
transition. Furthermore, even at the lowest temperature studied,
the transition appears quite broad, in particular when compared to
the metamagnetic transition in \srdzs , which consists of three
contributions each of them possessing a width of the order of a
tenth of a Tesla \cite{13,gegenwart}. We cannot exclude that
similar features are hidden in the broader peak in \caruoea \ but
one may note that the symmetry in \caruoea \ is orthorhombic
already at zero field. In \casrruo \ the mixed occupation of Ca
and Sr with distinct ionic radii induces strong intrinsic
disorder with local variations of the tilt and rotation angles
together with the concomitant local variation of the electronic
structure. Evidence for local disorder in the \casrruo -series
has recently been found in ARPES and STM studies
\cite{unordnung}. The intrinsic disorder most likely is
responsible for the broadening of the transition and it may
further  suppress any hysteresis. Keeping the strong microscopic
disorder in mind it appears very difficult to determine the
thermodynamic critical end-point in \caruoea \ or even to discuss
its existence at finite temperatures.

\begin{figure}
\centerline{\psfig{file=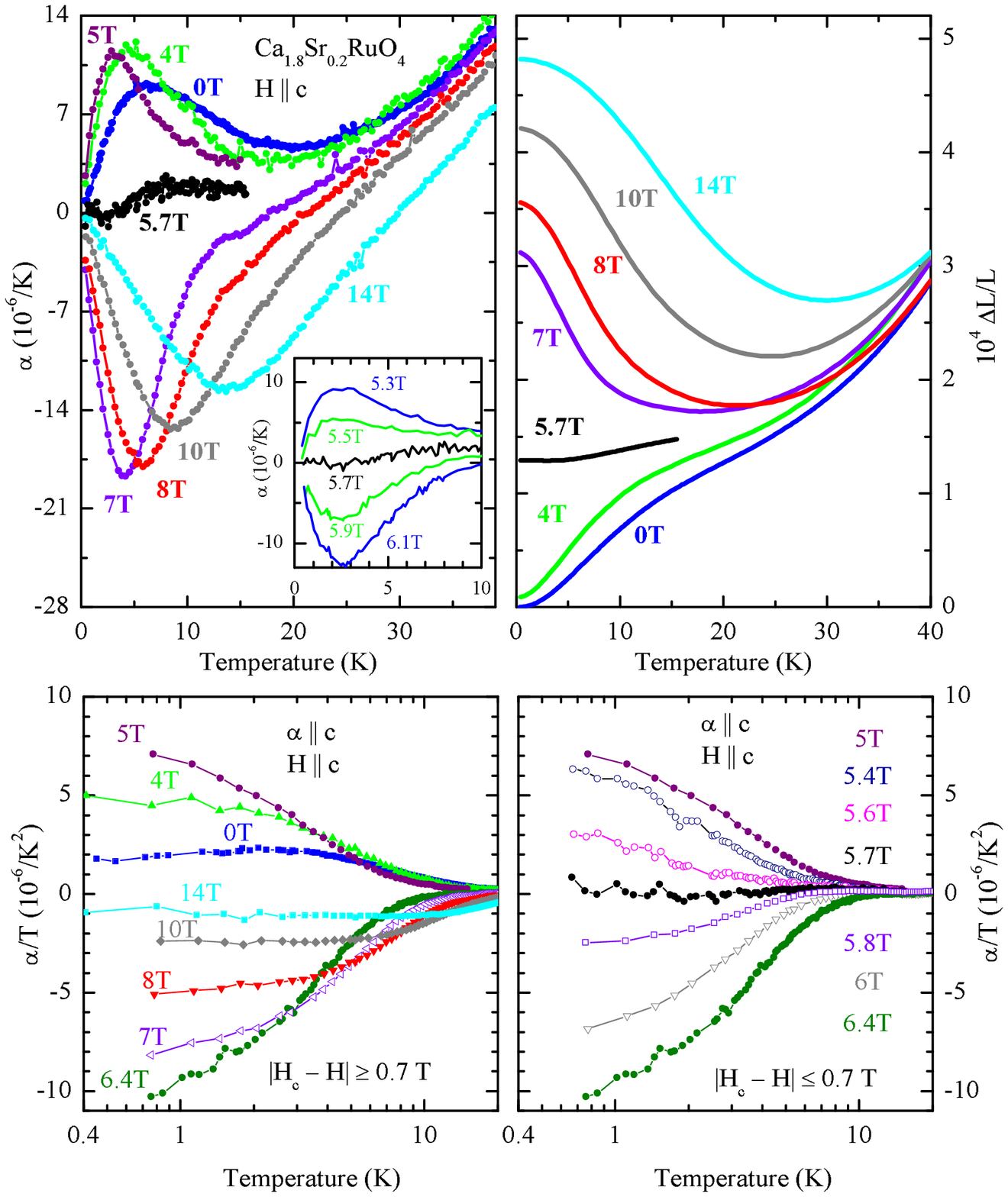,width=11cm}}
\caption{Thermal expansion of Ca$_{1.8}$Sr$_{0.2}$RuO$_4$
parallel to the $c$ axis in a longitudinal magnetic field. The
upper left panel shows the thermal expansion coefficient
$\alpha_c(T)=\frac{\partial c}{c\partial{\rm T}}$ and the upper
right panel the length change $\Delta c/c$ in fields below and
above the metamagnetic transition at $H_c \simeq 5.7$\,T. The
inset shows a magnified view of the $\alpha_c(T)$ anomaly in the
vicinity of the metamagnetic transition. In the lower panels,
$\alpha_c/T$ is plotted on a logarithmic scale in fields far away
from (left) and in the vicinity (right) of the metamagnetic
transition.} \label{fig:baier_fig2}
\end{figure}

Figure 4 shows the results of the thermal-expansion measurements
taken along the $c$-direction with the field parallel to $c$
(longitudinal configuration). Both the thermal expansion
coefficient $\alpha_c(T)=\frac{\partial{ c}}{c\partial{\rm T}}$
and the length change $\Delta c/c$ are shown in the upper panels.
One immediately recognizes that the strong thermal expansion
anomaly occurring around 20 \ K changes its sign upon increase of
the magnetic field in accordance with the idea that all these
effects are due to the orbital rearrangement \cite{kriener}.
Here, we want to discuss whether the effects across the
metamagnetic transition in \caruoea \ can be related with the
accumulation of entropy expected at the thermodynamic end-point.
For \srdzs \ it is argued that quantum criticality plays a
dominant role in spite of the first-order character of the
metamagnetic transition, since the end-point of the metamagnetic
transition would be sufficiently low in temperature
\cite{11,12,13}. It is therefore interesting to look for
signatures of a quantum-critical end-point in the thermodynamic
properties of \caruoea \ motivating us to extend the previous data
\cite{kriener} to lower temperatures. In the inset of the
upper-left panel of Figure 4 we show the thermal expansion
coefficient for magnetic fields close to the low-temperature
metamagnetic transition.  The anomalous effects seem to be
essentially suppressed when approaching the transition field.
This effect is also seen when plotting the ratio of the thermal
expansion coefficient over temperature $\alpha_c(T) \over T$,
see  the lower panels of Figure 4. Upon increase of the magnetic
field $\alpha_c(T) \over T$ first increases to exhibit maxima
slightly below the metamagnetic transition, see Figure 5. At the
transition it changes its sign and upon further field increase
there is a minimum slightly above the transition. The absolute
value of $\alpha_c(T) \over T$ is roughly symmetric in $H-H_{mm}$
and the distance of the two extrema is in agreement with the
width of the transition seen in the low-temperature
magnetostriction. Dividing the $\alpha_c(T) \over T$ values by
the analogous $C_p(T) \over T$ values one may determine the
Gr\"uneisen-parameter. The field dependence of the $C_p(T) \over
T$-ratio was reported in our previous paper \cite{kriener}. Upon
increasing the magnetic field at low temperature,  $C_p(T) \over
T$ increases only by about 20\% up to a maximum at the
metamagnetic transition and then drops rapidly above the
transition. Consequently, the Gr\"uneisen-parameter does not
diverge when approaching $H_{mm}$ in \caruoea .

\begin{figure}
\centerline{\psfig{file=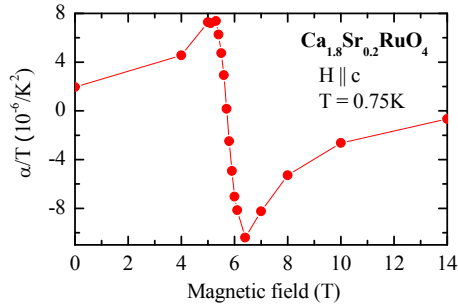,width=6cm}} \caption{
The $c$-axis thermal-expansion coefficient divided by the
temperature in \caruoea \ as a function of the magnetic field. }
\label{diffraction-c}
\end{figure}

From inelastic neutron scattering, it is known that \caruoea \
exhibits at least two magnetic instabilities
\cite{friedt-prl,steffens-unp} related with strongly enhanced
magnetic fluctuations. An incommensurate antoferromagnetic
contribution arising from  Fermi-surface effects appears to
compete with a quasi-ferromagnetic instability. The latter can be
directly deduced from the temperature dependence of the
macroscopic susceptibility \cite{friedt-prl,9}. For compositions
close to \caruoef \ a ferromagnetic cluster glass has even been
reported \cite{9}. At intermediate temperatures the macroscopic
susceptibility for $x=0.2$ exceeds that of \caruoef \ but it
exhibits a maximum at around 10\ K and is much smaller than that
for $x=0.5$ at low temperature. Compared to a Curie-Weiss
extrapolation the susceptibility for $x=0.2$ is significantly
reduced at the lowest temperatures. The incipient ferromagnetic
instability occurring in \caruoea \ as well as in \caruoef \
seems to get efficiently blocked through the structural anomaly
flattening the RuO$_6$-octahedron at low temperatures. This
effect may reduce the amplitude of the associated fluctuations
and enhance their characteristic energy. Through the transfer of
electrons into the \gbd , the ferromagnetic instability is
weakened possibly due to a shift of the van-Hove singularity. At
higher fields, the compound is forced into a ferromagnetic
ordering and, therefore,  quasi-ferromagnetic fluctuations are
weakened by further stabilizing this ferromagnetic ordering
explaining the reversed structural anomalies occurring upon
cooling for fields above $H_{mm}$. The sign change of the
thermal-expansion anomaly just at the transition field, its large
amplitude and its nearly symmetric behavior around the transition
field imply that the metamagnetic transition is related with
strong fluctuations \cite{garst}. The critical end-point of the
metamagnetic transition in \caruoea \ although hidden by the
intrinsic disorder of the system must be close within the
relevant energy scales.

\begin{figure}
\centerline{\psfig{file=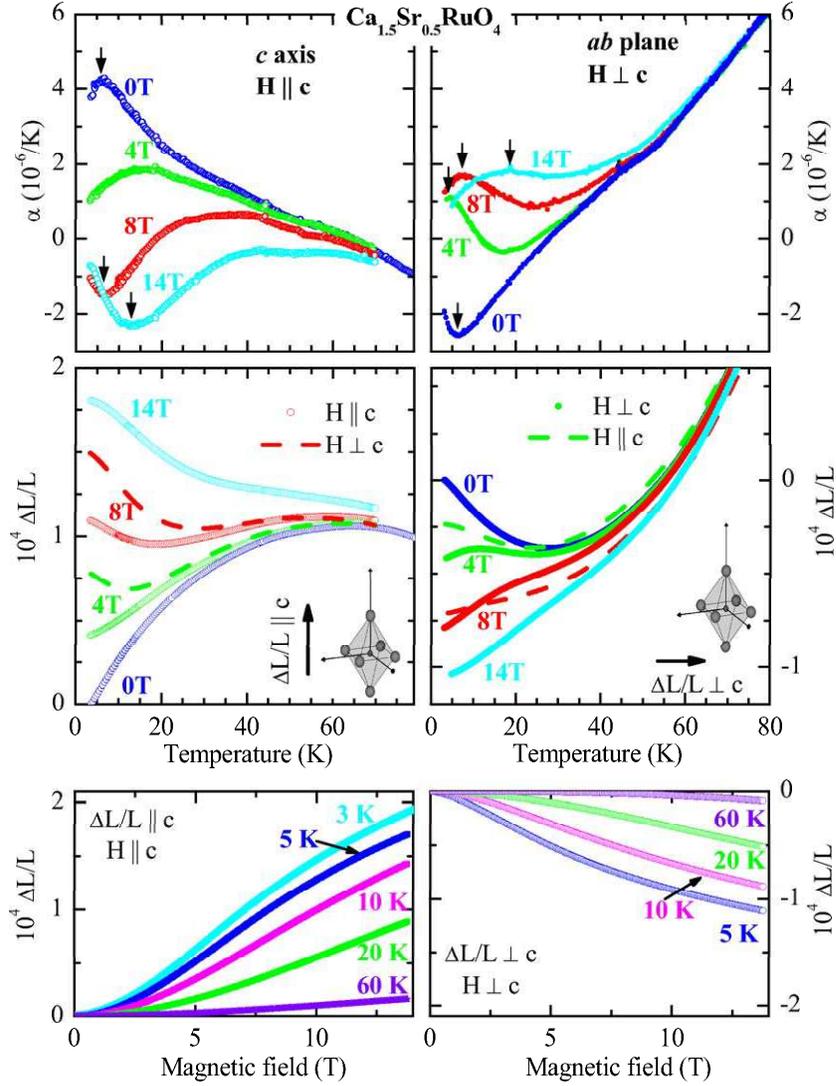,width=11cm}}
\caption{Thermal expansion and magnetostriction of
Ca$_{1.5}$Sr$_{0.5}$RuO$_4$ parallel (left) and perpendicular
(right) to the $c$ axis. The uppermost diagrams display
$\alpha(T)$ for both directions, each with longitudinal applied
magnetic field. The corresponding length change $\Delta L/L$ is
presented below. These diagrams show additionally the results
obtained from measurements in transverse magnetic field as broken
lines. The lowermost panels show the magnetostriction, each
recorded in longitudinal applied magnetic field.}
\label{fig:baier_fig4}
\end{figure}

Garst and Rosch \cite{garst} and Gegenwart et al. \cite{gegenwart}
have made quantitative predictions for a metamagnetic transition
related with a quantum-critical end-point which were already
tested for the \srdzs -compound. First $\alpha /T$ should vary as
$\vert H -H_{mm}\vert ^{-{4 \over 3}}$ at both sides of the
transition. The \caruoea -data shown in Fig. 5 clearly deviate
from such a behavior, in particular there is no divergence in the
experimental data. Close to the transition the microscopic
disorder may superpose positive and negative thermal expansion
anomalies cancelling each other. The almost complete suppression
of the anomaly close to the metamagnetic transition field is only
possible if the intrinsic ${\alpha \over /T}(H-H_{mm})$ dependence
is fully antisymmetric giving further weight to our interpretation
that the strongest fluctuations appear just at the metamagnetic
transition and that the critical end-point must be quite close.
These theories furthermore correctly predict that the thermal
expansion anomalies increase in temperature with increasing
$\vert H -H_{mm}\vert$. However, the scaling laws proposed for
the thermal expansion do not agree perfectly with our data
\cite{baier}. Again the intrinsic disorder might change the
temperature dependencies quite drastically. In addition the strong
antiferromagnetic fluctuations which are well established in
\casrruo \ will also interfere with the thermodynamic parameters.

\begin{figure}
\centerline{\psfig{file=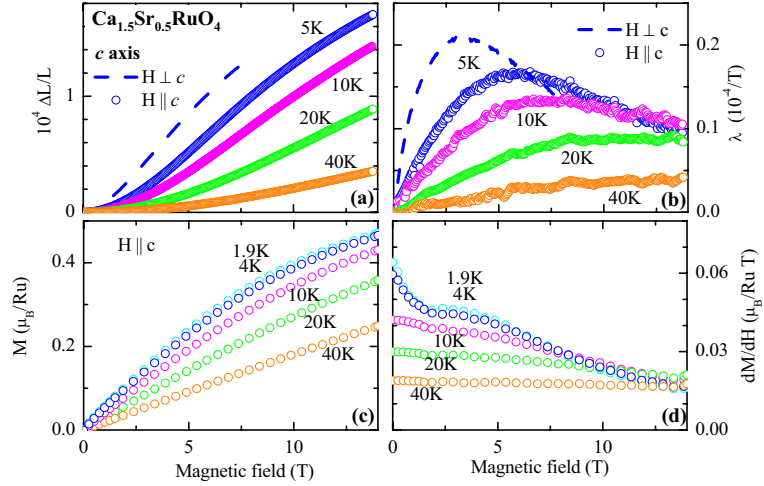,width=10cm}} \caption{
Comparison of the anomalous behavior of the $c$-axis
magnetostriction (upper panels) and the magnetization (lower
panels) for Ca$_{1.5}$Sr$_{0.5}$RuO$_4$. On the right, the data
obtained from measurements in a magnetic field applied along $c$
direction are shown. The diagrams on the left display the
corresponding derivative. The broken line in the upper panels
represent an example of the magnetostriction in a magnetic field
perpendicular to the $c$ axis.} \label{fig:baier_fig5}
\end{figure}

\subsection{ Magnetism in \caruoef }

Concerning the crystal structure, \caruoef \ differs from the
\caruoea -compound by the absence of the long-range tilt
distortion. This is visible in the  $c$-axis thermal-expansion
coefficient which is negative over a wide temperature interval,
see Figure 1. Due to the larger and positive thermal-expansion
coefficient in the $a,b$ plane the volume thermal expansion
however is positive also for $x=0.5$. Taking into account the
different thermal expansion behavior at intermediate
temperatures, the low-temperature $c$-axis anomalies are
qualitatively similar in \caruoea \ and in \caruoef , compare
Figures 6 and 4. This suggests that a metamagnetic transition
also occurs in \caruoef . However, all structural anomalies are
significantly smaller for $x=0.5$. The field-dependent thermal
expansion for $x=0.5$ was measured in the four configurations
with field and length change either parallel or perpendicular to
the $c$-axis. Again the difference in the longitudinal and
transverse configurations arise mainly from a shift in the
transition fields which is much smaller for fields oriented
perpendicular to the $c$-axis. In addition to the results
discussed for $x=0.2$, these \caruoef -data show that the
in-plane lattice parameters anomalously increase upon cooling in
zero field and become shorter in high fields. The
magnetostriction data shown in the lowest panels of Figure 6
confirm the opposite signs of the field-induced length changes
parallel and perpendicular to the $c$-axis. The volume
magnetostriction is about an order of magnitude smaller than the
uniaxial components confirming our interpretation that these
effects arise from an orbital rearrangement between the
$t_{2g}$-orbitals \cite{kriener}. The strong reduction of the
magnetostriction and of the thermal expansion anomalies for
$x=0.5$ must be related with the absence of the long-range tilt
deformation. Either the lattice in the crystal structure with a
simple rotational distortion is much harder thereby reducing the
structural response, or there is a direct interplay between the
tilt and electronic parameters. One may expect the tilt
deformation to act more strongly on the \dxz \ and \dyz \ orbitals
in contrast to the octahedron rotation which interferes mostly
with the \dxy \ orbitals \cite{5,6}.

Figure 7 compares the magnetization and the field-induced length
change along the $c$-direction and their derivatives versus
magnetic field. In the magnetization hysteresis cycles we find a
remanent magnetization of a few thousands of a \muB \ for fields
along and perpendicular to the $c$-direction in agreement with the
cluster-glass behavior reported in reference \cite{9}. The
underlying short-range ferromagnetic ordering seems to be another
consequence of the intrinsic disorder implied by the Ca/Sr mixing.
Even though the remanent magnetization is extremely small the
underlying ferromagnetic ordering has a strong impact on the
magnetization curves, in particular for the field within the
planes. At low field the magnetization sharply increases hiding
any metamagnetic transition at higher field. The magnetization
data shown in Figure 7 do not yield direct evidence for the
metamagnetic transition. Furthermore, the steep low-field
increase of the magnetization is not accompanied by the
magnetostriction in contrast to the close coupling between these
entities in \caruoea . The low-field increase of the
magnetization seems to fully arise from the short-range
ferromagnetic correlations; this feature is not related with the
metamagnetic transition and there is no comparable feature
observed for $x=0.2$. However, the field derivative of the
magnetization for \caruoef \ shown in the lower-right panel of
Figure 7 clearly exhibits two features: in addition to the finite
low-field value there is a clear shoulder at higher fields
resembling the peak at the metamagnetic transition in \caruoea .
This second feature corresponds to the maximum in the
magnetostriction (upper right panel). Therefore, we may conclude
that a metamagnetic transition in \caruoef \ still occurs,
although the associated jump in the magnetization is strongly
reduced compared to that in the \caruoea .

\section{Conclusions}

Detailed studies of the structural properties in \caruoea \ and in
\caruoef \ by diffraction and by dilatometer methods allow us to
clarify the microscopic and the thermodynamic aspects of the
metamagnetic transition in these materials. A temperature and
magnetic-field driven redistribution of the orbital occupation
seems to be responsible for the anomalous structural effects.
Upon cooling in zero field $3d$-electrons seem to move into the
\dxy \ orbitals causing a suppression of a quasi-ferromagnetic
instability. This effect is reversed either by cooling at high
magnetic field or by applying a magnetic field at low
temperature. The structural difference between \caruoea \ and
\caruoef \ consists in the long-range tilt deformation which is
found to strongly enhance the structural as well as the magnetic
anomalies. Even though the magnetization data in \caruoef \ do
not exhibit the well-defined metamagnetic transition, the field
derivative of the magnetization as well as the magnetostriction
clearly show that a qualitatively similar metamagnetic transition
also occurs in \caruoef . However, this material already exhibits
short-range ferromagnetic ordering at low temperature and zero
magnetic field which partially hides the metamagnetic transition.
The identical ferromagnetic instability is also present in
\caruoea \ at intermediate temperature, but here it is fully
suppressed at low temperature due to the octahedron tilting.

By analyzing the thermal expansion anomalies close to the
metamagnetic transition in \caruoea , we present evidence that the
related critical end-point must be close to the low-temperature
transition in the relevant scales. In particular, we find that the
$\alpha /T$-coefficient is nearly symmetric across the transition
field. The more precise scaling predictions for the thermal
expansion coefficient across a metamagnetic transition are,
however, not fulfilled in \caruoea .  In this material intrinsic
disorder as well as the competition of different magnetic
instabilities appear to  play an important role.

\section*{ACKNOWLEDGMENTS}
 We wish to dedicate this manuscript to Professor Hilbert von
 L\"ohneysen on the occasion of his 60th birthday. He has made
 numerous contributions to the field of strongly correlated
 electron systems and quantum phase transitions from which we all
 have greatly benefited.
This work was supported by the DFG through the
Sonderforschungsbereich 608.

\end{document}